\documentclass[twocolumn,epjc3]{svjour3}  
\smartqed  % flush right qed marks, e.g. at end of proof
\RequirePackage{graphicx}
\usepackage{tikz}
\RequirePackage[colorlinks,citecolor=blue,urlcolor=blue,linkcolor=blue]{hyperref}
\journalname{Eur. Phys. J. C}

\RequirePackage{lineno}\newdimen\linenumbersep\linenumbersep=2pt
\usepackage[T1]{fontenc}
\usepackage[utf8]{inputenc}
\usepackage{subfigure}
\usepackage{mathtools}
\usepackage{longtable}
\usepackage{lineno}
\usepackage{array, booktabs, color, eurosym, longtable, pbox, pdfpages, siunitx, url, verbatim}

\usepackage{xspace}
\usepackage{wrapfig}
\usepackage{multirow}
\usepackage{blindtext}

\usepackage{babel}
\usepackage{booktabs}
\newcommand{\Ricochet}{\textsc{Ricochet}}

\makeatletter
\let\cl@chapter\undefined
\makeatletter
\usepackage[capitalise]{cleveref}

\usepackage{booktabs}
\usepackage{multirow}
\usepackage{textgreek}
\usepackage{siunitx}
\usepackage{xspace}

\makeatletter

\begin{document}

\title{First demonstration of 30 eVee ionization energy resolution with \Ricochet{} germanium cryogenic bolometers}

%
% need to make sure the affiliations get sorted / listed in a way
% that isn't too awkward
%
\author{
{C.~Augier}\thanksref{a}\and
{G.~Baulieu}\thanksref{a}\and
{V.~Belov}\thanksref{h}\and
{L.~Berg\'e}\thanksref{b}\and
{J.~Billard}\thanksref{a,e1}\and
{G.~Bres}\thanksref{g}\and
{J-.L.~Bret}\thanksref{g}\and
{A.~Broniatowski}\thanksref{b}\and
{M.~Calvo}\thanksref{g}\and
{A.~Cazes}\thanksref{a}\and
{D.~Chaize}\thanksref{a}\and
{M.~Chala}\thanksref{c}\and
{M.~Chapellier}\thanksref{b}\and
{L.~Chaplinsky}\thanksref{f}\and
{G.~Chemin}\thanksref{c}\and
{R.~Chen}\thanksref{d}\and
{J.~Colas}\thanksref{a,e2}\and
{E.~Cudmore}\thanksref{j}\and
{M.~De Jesus}\thanksref{a}\and
{P.~de Marcillac}\thanksref{b}\and
{L.~Dumoulin}\thanksref{b}\and
{O.~Exshaw}\thanksref{g}\and
{S.~Ferriol}\thanksref{a}\and
{E.~Figueroa-Feliciano}\thanksref{d}\and
{J.-B.~Filippini}\thanksref{a}\and
{J.~A.~Formaggio}\thanksref{e}\and
{S.~Fuard}\thanksref{k}\and
{J.~Gascon}\thanksref{a}\and
{A.~Giuliani}\thanksref{b}\and
{J.~Goupy}\thanksref{g}\and
{C.~Goy}\thanksref{c}\and
{C.~Guerin}\thanksref{a}\and
{E.~Guy}\thanksref{a}\and
{P.~Harrington}\thanksref{e}\and
{S.~A.~Hertel}\thanksref{f}\and
{M.~Heusch}\thanksref{c}\and
%{C.~F.~Hirjibehedin}\thanksref{i}\and
{Z.~Hong}\thanksref{j}\and
{J.-C.~Ianigro}\thanksref{a}\and
{Y.~Jin}\thanksref{l}\and
{A.~Juillard}\thanksref{a,e3}\and
{D.~Karaivanov}\thanksref{h}\and
{S.~Kazarcev}\thanksref{h}\and
{J.~Lamblin}\thanksref{c}\and
{H.~Lattaud}\thanksref{a}\and
{M.~Li}\thanksref{e}\and
{A.~Lubashevskiy}\thanksref{h,e6}\and
{S.~Marnieros}\thanksref{b}\and
{N.~Martini}\thanksref{a,e4}\and
{D.~W.~Mayer}\thanksref{e}\and
{J.~Minet}\thanksref{g}\and
{A.~Monfardini}\thanksref{g}\and
{F.~Mounier}\thanksref{a}\and
{V.~Novati}\thanksref{d}\and
%{W.~D.~Oliver}\thanksref{i}\and
{E.~Olivieri}\thanksref{b}\and
{C.~Oriol}\thanksref{b}\and
{L.~Ovalle Mateo}\thanksref{d}\and
{P. K.~Patel}\thanksref{f}\and
{E.~Perbet}\thanksref{c}\and
{H.~D.~Pinckney}\thanksref{f}\and
{D.~V.~Poda}\thanksref{b}\and
{D.~Ponomarev}\thanksref{h,e6}\and
{F.~Rarbi}\thanksref{c}\and
{J.-S.~Real}\thanksref{c}\and
{T.~Redon}\thanksref{b}\and
{F.~C.~Reyes}\thanksref{e}\and
{A.~Robert}\thanksref{k}\and
{S.~Rozov}\thanksref{h}\and
{I.~Rozova}\thanksref{h}\and
{S.~Scorza}\thanksref{c}\and
{B.~Schmidt}\thanksref{d,e5}\and
{Ye.~Shevchik}\thanksref{h}\and
{T.~Soldner}\thanksref{k}\and
{J.~Stachurska}\thanksref{e}\and
{A.~Stutz}\thanksref{c}\and
{L.~Vagneron}\thanksref{a}\and
{W.~Van De Pontseele}\thanksref{e}\and
{F.~Vezzu}\thanksref{c}\and
%{S.~Weber}\thanksref{i}\and
{L.~Winslow}\thanksref{e}\and
{E.~Yakushev}\thanksref{h}\and
{D.~Zinatulina}\thanksref{h} 
 -- the \Ricochet{} Collaboration
}

\institute{{Univ Lyon, Université Lyon 1, CNRS/IN2P3, IP2I-Lyon, F-69622 Villeurbanne, France}\label{a}\and
{Department of Nuclear Spectroscopy and Radiochemistry, Laboratory of Nuclear Problems, JINR, Dubna, Moscow Region, Russia 141980}\label{h}\and
{Université Paris-Saclay, CNRS/IN2P3, IJCLab, 91405 Orsay, France}\label{b}\and
{Univ. Grenoble Alpes, CNRS, Grenoble INP, Institut Néel, 38000 Grenoble, France}\label{g}\and
{Univ. Grenoble Alpes, CNRS, Grenoble INP, LPSC-IN2P3, 38000 Grenoble, France}\label{c}\and
{Department of Physics, University of Massachusetts at Amherst, Amherst, MA, USA 01003}\label{f}\and
{Department of Physics and Astronomy, Northwestern University, IL, USA}\label{d}\and
{Department of Physics, University of Toronto, ON, Canada M5S 1A7}\label{j}\and
{Laboratory for Nuclear Science, Massachusetts Institute of Technology, Cambridge, MA, USA 02139}\label{e}\and
{Institut Laue Langevin, 38042 Grenoble, France }\label{k}\and
%{Lincoln Laboratory, Lexington, MA, USA}\label{i}\and
{Universit\'e Paris-Saclay, CNRS, C2N, 91120 Palaiseau, France}\label{l}
}
%Email for the corresponding author

\thankstext{e1}{e-mail: j.billard@ipnl.in2p3.fr}
\thankstext{e2}{e-mail: j.colas@ip2i.in2p3.fr}
\thankstext{e3}{e-mail: a.juillard@ipnl.in2p3.fr}
\thankstext{e4}{e-mail: n.martini@ip2i.in2p3.fr}
\thankstext{e5}{Currently at CEA, Gif-sur-Yvette, France}
\thankstext{e6}{Also at LPI RAS, Moscow, Russia}

\maketitle

\begin{abstract}
The future \Ricochet{} experiment aims to search for new physics in the electroweak sector by measuring the Coherent Elastic Neutrino-Nucleus Scattering process from reactor antineutrinos with high precision down to the sub-100 eV nuclear recoil energy range. While the \Ricochet{} collaboration is currently building the experimental setup at the reactor site, it is also finalizing the cryogenic detector arrays that will be integrated into the cryostat at the Institut Laue Langevin in early 2024. In this paper, we report on recent progress from the Ge cryogenic detector technology, called the CryoCube. More specifically, we present the first demonstration of a 30~eVee (electron equivalent) baseline ionization resolution (RMS) achieved with an early design of the detector assembly and its dedicated High Electron Mobility Transistor (HEMT) based front-end electronics. This represents an order of magnitude improvement over the best ionization resolutions obtained on similar heat-and-ionization germanium cryogenic detectors from the EDELWEISS and SuperCDMS dark matter experiments, and a factor of three improvement compared to the first fully-cryogenic HEMT-based preamplifier coupled to a CDMS-II germanium detector. Additionally, we discuss the implications of these results in the context of the future \Ricochet{} experiment and its expected background mitigation performance.

\end{abstract}

%\flushbottom

\section{Introduction}

The recent first observation of Coherent Elastic Neutrino-Nucleus Scattering (CENNS) by the COHERENT collaboration has opened new avenues to search for physics beyond the standard model~\cite{Akimov:2017ade,COHERENT:2020iec}.  The \Ricochet{} collaboration~\cite{Ricochet:2022pzj,Ricochet:2021rjo} is aiming for a percentage-level CENNS measurement down to sub-100 eV nuclear recoil energies where signatures of such new physics may arise~\cite{Billard:2018jnl}. These include for instance the existence of sterile neutrinos and of new mediators that could be related to the long lasting dark matter problem, and the possibility of non-standard neutrino   interactions that would dramatically affect our understanding of the electroweak sector. 

The future \Ricochet{} experiment will be deployed at the Institut Laue Langevin (ILL) within the H7 experimental site~\cite{Ricochet:2022pzj} to measure with high precision the CENNS process from reactor antineutrinos. To achieve its goal, the \Ricochet{} experiment seeks to utilize a kg-scale cryogenic detector payload combining the CryoCube and the Q-Array, two cryogenic detector technologies consisting of 18-to-27 Ge and 9 Zn crystals, respectively. The cryogenic detectors will be located 8.8~m away from the 58~MW nominal thermal power reactor leading to a CENNS event rate of approximately 12.8 and 11.2~events/kg/day with a 50~eV energy threshold in the Ge and Zn target crystals, respectively. Both technologies are being optimized to combine a sub-100~eV energy threshold with particle identification capabilities to reject both the gamma-induced electronic recoil background, and
the low energy excess observed in all low-energy threshold cryogenic detectors~\cite{Fuss:2022fxe}. Above the threshold of particle identification, the neutron-induced nuclear recoils are  expected to be the limiting background to the future \Ricochet{} experiment. A recent characterization of this neutron background has shown that, despite the 15~m.w.e artificial overburden at the ILL-H7 site, the \Ricochet{} neutron background will be dominated by cosmogenic neutrons and that \Ricochet{}'s expected CENNS signal-to-noise ratio should be around unity~\cite{Ricochet:2022pzj}.

 \begin{figure*}[t]
\begin{center}
\includegraphics[width=\textwidth,angle=0]{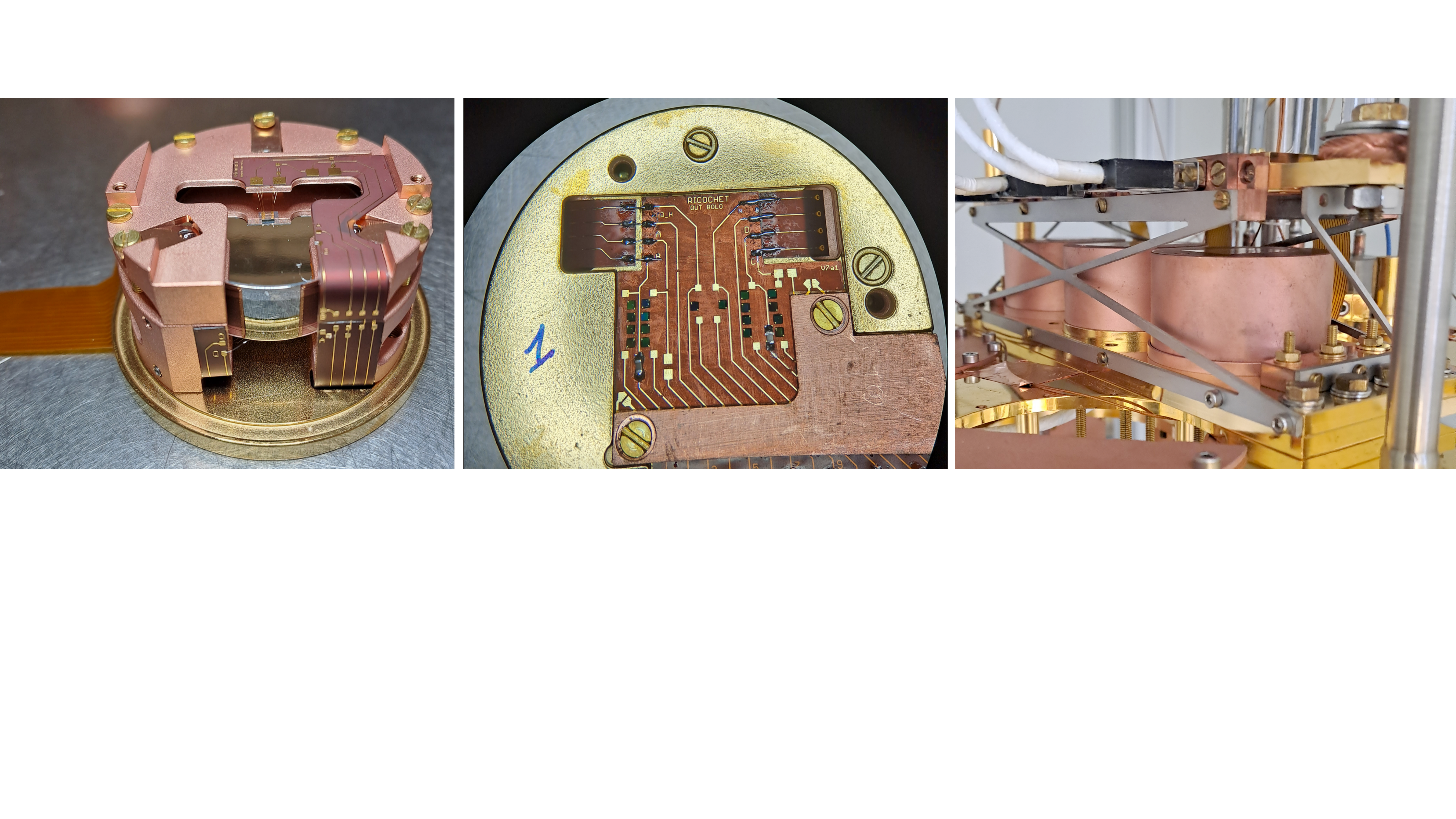}
\vspace{-4.2cm}
\caption{{\bf Left:} Photo of the RED177 planar detector. The heat and ionization signals are respectively measured with an NTD sensor, glued in the middle of the top surface, and top and bottom aluminum electrodes. The electrical connection between the 10~mK and 1~K stages is done via a 5~cm long and 100~$\mu$m thick flexible Kapton PCB with constantan leads visible on the left-hand side of the photo. {\bf Middle:} Photo of the 10~mK stage electronics mounted on the bottom of the detector holder. The latter hosts the load and feedback resistors, as well as the coupling capacitors. {\bf Right:} MiniCryoCube assembly, hosting the detectors RED177, RED227, and RED237 (from right to left), mounted on the mixing chamber stage of the IP2I R\&D cryostat. The 1K stage of the MiniCryoCube lies above the three detectors and houses the HEMT preamplifiers. This stage is thermally anchored to the 1~K plate via a cold finger. } 
\label{fig:Photo}
\end{center}
\end{figure*}

Similarly to the EDELWEISS experiment~\cite{Armengaud:2017rzu}, particle identification with the CryoCube will be achieved thanks to the simultaneous measurement of heat and ionization energies. The Ge crystals are equipped with a germanium Neutron Transmutation Dopped (NTD) heat sensor and aluminum electrodes to collect signals from the ionization induced electron-hole pairs. Setting aside the low energy excess (see Sec.~\ref{sec:HO} for a dedicated discussion), to achieve the targeted  CENNS sensitivity of the \Ricochet{} experiment, the CryoCube heat and ionization channels are designed to achieve 10~eV and 20~eVee (eV electron equivalent) baseline energy resolutions (RMS), respectively~\cite{RICOCHET:2021gkf}. With such performance, the CryoCube discrimination energy threshold is expected to be on the order of 100~eV, allowing for a percentage precision CENNS measurement. While both the SuperCDMS~\cite{SuperCDMS:2020aus} and EDELWEISS~\cite{EDELWEISS:2019vjv} collaborations have already demonstrated heat energy resolutions on the $\mathcal{O}$(10)~eV-scale with tens of gram target crystals, state-of-the-art ionization energy resolution would be the limiting factor for the CryoCube. At present, ionization resolutions of 220 eVee and 350 eVee, for EDELWEISS~\cite{Hehn:2016nll} and SuperCDMS~\cite{SuperCDMS:2014cds} respectively, leading to keV-scale discrimination energy thresholds, are an order of magnitude greater than our \Ricochet{} design requirement. It is worth mentioning that baseline ionization energy resolutions on the order of 30~eVee have already been achieved in large p-type point contact (PPC) Ge detectors operated at 77~K~\cite{CDEX:2017hbo,nGeN:2022uje,Collar:2021fcl,CONUS:2020skt}. This was accomplished in part by minimizing the input capacitance to the preamplifier stage of their silicon Field Effect Transistors (FETs). The lower capacitance is achieved by 1) reducing the capacitance of the electrode itself, and 2) reducing the capacitance of the cabling between the FET and the electrode by reducing the distance between the two. The reduction of cabling capacitance is much more difficult to achieve with sub-100~mK cyogenic detectors with FET-based electronics operated at 100~K. Indeed, reducing the distance between the detectors and the FET to a few centimeters would impose an overwhelmingly large heat load to both the cryostat and the detectors preventing any simultaneous heat measurement. Additionally, the highly asymmetric design of the PPC electrodes produces inhomogeneities in the electric field that would strongly affect the charge collection efficiency in the Ge crystal operated at 10~mK with low electrode bias voltages ($\leq$5~V/cm), hence forbidding any particle discrimination capabilities.

Thanks to their much lower intrinsic current noise, input capacitance and working temperature allowing for reduced cabling length and related stray capacitance, High Electron Mobility Transistors (HEMT), developed by the Center for Nanoscience and Nanotechnology (C2N) and commercialized by CryoHEMT~\cite{articleHEMT}, have been identified as a replacement for the standard FET to improve the ionization resolution of semiconducting cryogenic bolometers. A first fully cryogenic HEMT-based preamplifier has been successfully operated with a CDMS-II Ge detector and achieved an unprecedented 91~eVee (RMS) ionization resolution~\cite{Phipps:2016gdx}. This result, while falling short of the \Ricochet{} goals (see Sec.~\ref{sec:HO}), demonstrated the interest of using HEMT-based preamplifiers in the context of cryogenic detectors, and paved the way to the ongoing CryoCube developments. 

In this paper we report on the first demonstration of 30~eVee (RMS) ionization baseline energy resolution with \Ricochet{} Ge cryogenic detectors operated at 15~mK. This result has been achieved with an early design of a sub-element of the CryoCube detector array, called MiniCryoCube, hosting three 38~g Ge detectors with their dedicated HEMT-based front-end electronics thermally anchored at 1~K. The paper is organized as follows: in Sec.~\ref{sec:setup} we present the experimental setup including the CryoCube holding structure and the Ge detector design; in Sec.~\ref{sec:HEMTsc} we detail the HEMT-based preamplifier; in Sec.~\ref{sec:processing} and Sec.~\ref{sec:results} we present our data processing and results, respectively. Finally, we discuss the implication of these first results in the context of the low-energy excess mitigation in Sec.~\ref{sec:HO}, and give our conclusions in Sec.~\ref{sec:conclusion}.

\section{Experimental setup}
\label{sec:setup}

The first tests of charge preamplification with HEMTs presented hereafter were performed with three CryoCube prototype detectors: RED177, RED227, and RED237. Each detector consists of a 38~g Ge cylindrical crystal of 30~mm diameter and 10~mm height equipped with two planar electrodes and a 2$\times$2~mm$^2$ NTD sensor\footnote{The NTD sensors used for the CryoCube detector array have been retrieved from the EDELWEISS-III experiment~\cite{EDELWEISS:2016nzl} and modified to match the CryoCube specifications.} glued in the middle of the top electrode (see~\cite{RICOCHET:2021gkf,misiak:tel-03328713} for more details about the CryoCube detector designs).

Figure~\ref{fig:Photo} (left panel) shows a photo of the RED177 planar detector with its Ge crystal held in its copper casing. Though not read out in this work, the NTD sensor is both electrically and thermally connected to the flexible Kapton printed-circuit board (PCB) thanks to a total of six 25~$\mu$m diameter gold wirebonds. The aluminum electrodes fully cover the flat top and bottom surfaces. In order to limit charge trapping on the lateral side, each electrode extends on that surface by 2~mm, hence limiting the height of the exposed bare Ge surface to 6~mm. The electrodes are electrically connected to the flexible PCB thanks to a total of eight 25~$\mu$m diameter aluminum wirebonds. The detector holder has been designed to both minimize vibration induced noise and reduce electrical capacitance by ensuring a minimal distance between the aluminum electrodes and the copper casing of 3~mm~\cite{RICOCHET:2021gkf}. From our simulations we expect a mutual capacitance between each electrode and the copper casing of 4.06~pF, and a mutual capacitance between the two electrodes of 10.86~pF~\cite{misiak:tel-03328713}. 

The middle panel of Fig.~\ref{fig:Photo} is a picture of the gold plated circular copper piece mounted on the bottom of the detector holder. It hosts the feedback and load resistors as well as the coupling capacitors (see Sec.~\ref{sec:HEMTsc}), heat sinking them to 10~mK. Lastly, the detector and its 10~mK stage electronic components are electrically connected to their dedicated 1~K front-end electronics via a 100~$\mu$m thick flexible Kapton PCB with unshielded constantan leads (see Fig.~\ref{fig:Photo} left panel).

Figure~\ref{fig:Photo} (right panel) shows a fully integrated MiniCryoCube array hosting three planar detectors, namely RED177, RED227, and RED237, mounted on the 10~mK mixing chamber stage of the R\&D cryostat at the Institute of Physics of the 2 Infinities (IP2I) in Lyon, France. The upper 1~K stage of the MiniCryoCube, only 5~cm above its lower 10~mK stage, is thermally anchored to the 1~K still plate thanks to a 15~mm diameter copper rod and a copper braid. The two stages of the MiniCryoCube are mechanically coupled to each other thanks to 1.6 mm thick X- and Z-shaped titanium alloy Ti-15-3-3-3~\cite{Timetal} support structures on the front/back and sides, respectively. With a set of heaters and thermometers located on the two MiniCryoCube stages,  we measured that the total heat load from the 1~K stage on the 10~mK stage, only 5~cm below, is about 1~$\mu$W. With a mixing chamber temperature set to 15~mK, the temperature at the detectors was found to be about 16~mK. Such thermal performance allows for an optimal operation of both the heat and ionization channels of all three planar detectors simultaneously. Lastly, each HEMT-based 1~K preamplifier PCB is terminated by a 25-pin micro-D connector which interfaces with the readout cabling going directly to a 300~K feedthrough. To quantify the impact of the cabling impedance on the charge amplifier performance, all three detectors were connected via stainless steel shielded coaxial cables with either low-impedance (5~\textOmega{}) Cu leads (RED177 and RED227) or  high-impedance (125~\textOmega{}) constantan leads (RED237). 

The room temperature electronics consists of a BiLT-BN103 chassis equipped with low-noise DC sources BE2142 from iTEST and custom filters. The latter are used to bias the HEMT and the detector's electrodes. The output signal is fed to a SR5184 amplifier from Signal Recovery, and low-pass filtered with a 8-order Bessel filter from KEMO with a cut-off frequency of 63 kHz. The signal streams are digitized at 100 or 200 kHz with a 16-bit National Instrument NI-6218 DAQ, and stored on disk for further data processing and analysis.

Prior to their installation in the cryostat the Ge detectors were exposed to an AmBe source, emitting $6\times 10^5$ neutrons per second, for about 114~h (RED177 and RED227) and 15~h (RED237) for $^{71}$Ge activation. This activation from neutron capture will lead to electron-capture decays from the K/L/M shells producing X-ray lines with summed energies of 10.37~keV, 1.30~keV, and 160~eV, respectively (see~\cite{EDELWEISS:2020fxc}). On top of being monoenergetic, these decays are uniformly distributed throughout the entire Ge crystal volume,  allowing for a precise study of the detector response and calibration.

\section{HEMT based common source preamplifier}
\label{sec:HEMTsc}

\begin{figure*}[t]
    \centering
    \includegraphics[width=\textwidth]{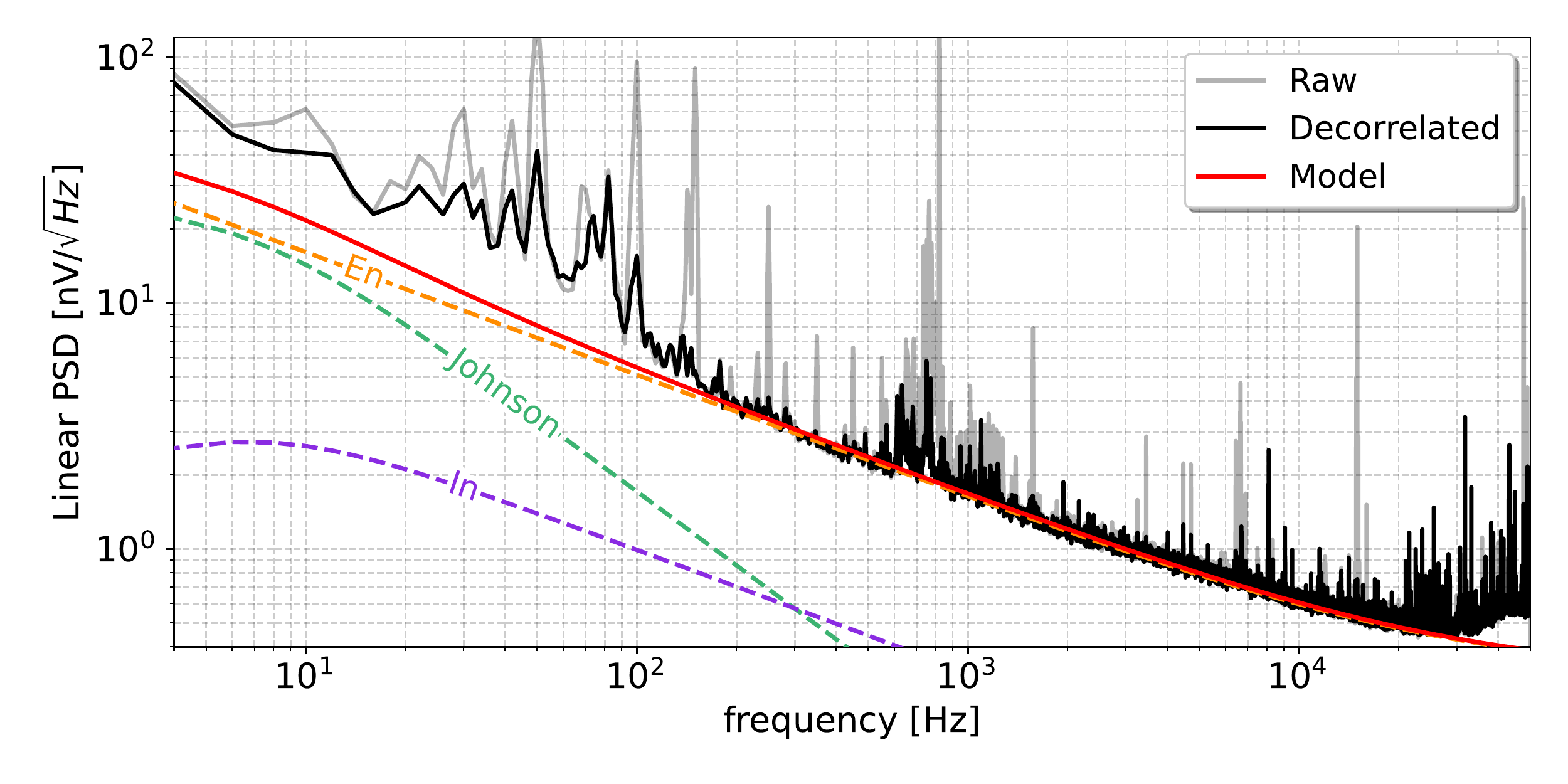} %{no_signal_fit.pdf} %o_signal_fit.pdf
    \vspace{-0.8cm}
    \caption{Differential ($V_B - V_A$) noise power spectrum of RED227 with $V_{ds}$ = 100 mV and $I_{ds}$ = 300 $\mu$A and the mixing chamber at 17~mK. The black, grey and red solid lines show the decorrelated and raw data, and our noise model considering a parasitic capacitance of 20~pF, respectively. The total contributions from the two ionization channels A and B of the current ($I_n$), voltage ($E_n$), and Johnson noise sources are also shown as dashed purple, orange, and green lines, respectively.}
    \label{fig:PSDRED177}
\end{figure*}

The three HEMT-based preamplifiers considered in this work are  in a standard common source
configuration with a 1~k\textOmega{} load resistor on the drain ($R_d$)~\cite{Juillard:2019njs}. The choice of using a common source scheme comes from its wide range of application. The latter is well suited to study the intrinsic properties of the transistor ({\it e.g.} bias curves, gain, and voltage noise), and leads to reasonable output gain of about 10 which is enough to be coupled to low-noise room temperature amplifiers~\cite{Ricochet:2021cqv}. The HEMTs and drain resistors are mounted on a custom PCB encapsulated in a copper chassis on the MiniCryoCube 1~K stage. Following our HEMT-based preamplifier model optimization~\cite{Juillard:2019njs}, we only used 4.6~pF input capacitance HEMTs from CryoHEMT~\cite{articleHEMT}. The dissipated power from each readout channel is given by:

\begin{equation}
    P = V_{ds}I_{ds} + R_dI_{ds}^2,
    \label{eq:Pdis}
\end{equation}
with $V_{ds}$ and $I_{ds}$ the drain-source voltage and current bias, respectively. The standard HEMT bias operating point considered in the following of this work is \{$V_{ds}$ = 100 mV, $I_{ds}$ = 300 $\mu$A\} leading to a dissipation power per HEMT channel of 120 $\mu$W. In comparison to \cite{Phipps:2016gdx}, our preamplifier design only uses one HEMT per ionization channel leading to significantly reduced heat load from 1~mW to 120~$\mu$W per channel. With a maximum of 108 ionization channels in total for a fully instrumented CryoCube with 27 detectors, each equipped with up to 4 ionization channels~\cite{RICOCHET:2021gkf,misiak:tel-03328713}, we expect a total heat load of about 13~mW on the still stage, which can be handled by the \Ricochet{} cryostat.

The 1~K HEMT-based common source preamplifier is connected to the 10~mK stage electronics which are located underneath the detector holder (see Fig.~\ref{fig:Photo} middle panel). This stage encapsulates the 800~M\textOmega{} total load and feedback resistors from Mini Systems Inc., and the 2~nF coupling capacitors. These components are glued on a 100~$\mu$m thick flexible Kapton PCB, with copper leads, that goes around the copper casing to connect both the heat and ionization sensors to the readout electronics. According to capacitive measurements performed at room temperature using  a precision RLC meter QuadTech 7600B and COMSOL\textsuperscript{\textregistered} simulations, we found that the cabling parasitic capacitance from this flexible Kapton PCB should be between 15~pF and 25~pF, hence comparable to the detector's mutual capacitances. According to our model, with this prototype cabling, we expect to achieve baseline energy resolutions between 27~eVee and 35~eVee (RMS) considering a differential readout and data processing (see  Sec.~\ref{sec:processing}). Experimental efforts are onging to further reduce the parasitic capacitance from the flexible Kapton PCB down to only a few pF in order to achieve the targeted 20~eVee (RMS) baseline ionization resolution (see Sec.~\ref{sec:conclusion}).

\section{Data processing}
\label{sec:processing}

 \begin{figure*}[t]
\begin{center}
\includegraphics[width=\textwidth,angle=0]{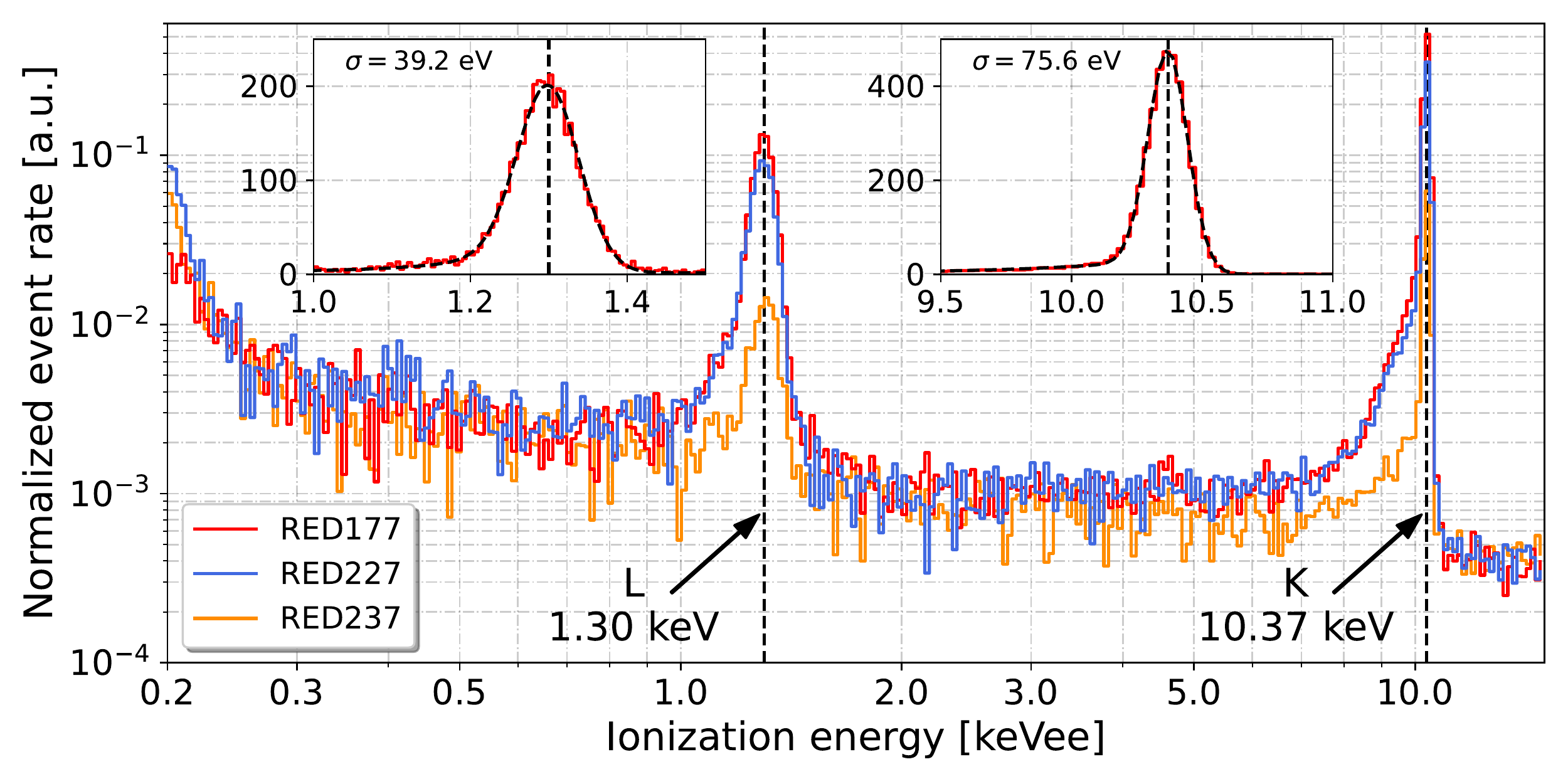}
\caption{Observed unit normalized, between 0 and 15 keV, energy spectra from RED177 (red), RED227 (blue), and RED237 (orange). The three detectors were operated with the two MiniCryoCube stages at 17~mK and 1.135~K, respectively. The data were acquired continuously for 92~h (RED177), 75~h (RED227), and 65~h (RED237) with the HEMT biased at $V_{ds}$ = 100 mV and $I_{ds}$ = 300 $\mu$A. The inset figures show, for RED177, the fit (black dashed line) to the K- and L-electron capture decay X-ray lines from the activated $^{71}$Ge at 1.30~keV and 10.37~keV (see text for details). Note that the binning along the x-axis of the normalized energy spectra is logarithmic, and linear for the inset figures.} 
\label{fig:spectra}
\vspace{-0.5cm}
\end{center}
\end{figure*}

Following each data acquisition, the data streams from the two ionization channels A (top electrode) and B (bottom electrode) are processed using the pipeline detailed in~\cite{Colas:2021pxr}. As the parasitic capacitance from the cabling is similar to that of the detector's electrodes the internal noise sources from the HEMT preamplifier can propagate from one channel to the other. This results in some non-negligible fraction of the noise to be correlated between the two channels A and B, in addition to any correlated electromagnetic pick-up noise sources from the environment. As one electrode collects the holes and the other the electrons, the difference $V_B - V_A$ contains the summed signal amplitudes while the sum $V_B + V_A$ contains most of the observed noise from the two channels. The signals are therefore decorellated using an algorithm inspired by~\cite{Allen:1999wh} about common noise sources subtraction. The result of this decorrelation is shown in Fig.~\ref{fig:PSDRED177}, where the gray and black solid lines compare the noise power spectra of $V_B - V_A$ for RED227 before and after this decorellation procedure. 

The red line on Fig.~\ref{fig:PSDRED177} is the prediction of our noise model adapted from~\cite{Juillard:2019njs} assuming a parasitic capacitance of 20~pF. The contributions of the total current, voltage and Johnson noise sources from the two ionization channels are also shown. Though a detailed study about the comparison between our HEMT-based preamplifier noise modelisation and the data is ongoing, we can already notice that our model is in reasonable agreement with the data, especially between 100~Hz and 30~kHz. We found that using this newly developed differential decorrelation processing method we improved the baseline energy resolutions by  10-to-20\%. For the rest of this work we will only consider the decorrelated differential signal.

\section{Results}
\label{sec:results}
 
In order to study the response of the three detectors and of these newly developed HEMT-based preamplifiers, continuous streams of 92, 75 and 65 hours of data were accumulated on the detectors RED177, RED227 and RED237, respectively, from the 26th of January to the 5th of February 2023. The A and B electrodes were respectively biased at $+$2~V and $-$2~V, and the detector and preamplifer stages of the MiniCryoCube were stable at 17~mK and 1.135~K, respectively.

The resulting energy spectra, normalized to unity over 0 and 15~keV, from all three detectors are shown in Fig.~\ref{fig:spectra}. Each spectrum has been obtained after applying cuts to reject pile-up events and saturated traces. From Fig.~\ref{fig:spectra} one can clearly see the X-ray lines at 1.30 and 10.37 keV emitted in the electron capture of $^{71}$Ge. These lines exhibit a non-Gaussian shoulder towards lower energies. This well-known feature is due to incomplete charge collection, and was quantified in~\cite{RICOCHET:2021gkf} where we found an upper limit on the fraction of incomplete charge collection events of about 10\% for the planar detector design. 
The experimental resolutions at 1.30 and 10.37~keV listed in Table~\ref{tab:fano} are the standard deviations of the Gaussian component obtained in least-square fits such as those shown in Fig.~\ref{fig:spectra} (see inset panels), where the peaks are modeled as a Gaussian combined with an exponential tail. The comparison of these resolutions with the achieved baselines, given in the first column of Table~\ref{tab:fano}, yields Fano factor~\cite{PhysRev.70.44} values of $0.145\pm 0.011$ and $0.1528\pm 0.0012$ at 1.30 and 10.37 keV, respectively. These derived Fano factor values, computed assuming a mean energy to create an electron-hole pair of 3.0~eV~\cite{EDELWEISS:2006bcu,knoll2010radiation} and using the weighted means of all three detectors\footnote{Note that for the Fano factor estimation at 10.37~keV the inconsistent result from RED237 was not used in the weighted average.} with statistical error only, are significantly higher than its standard value  of $F = 0.1057\pm 0.0002$ at 5.9~keV for Ge detectors operated at 77~K~\cite{LOWE1997354}. This could be explained by both the incomplete charge collection and the event-to-event gain variations expected with these common source preamplifiers which are not in a closed loop configuration~\cite{Ricochet:2021cqv}.  In addition to the $^{71}$Ge K- and L-shell electron capture lines, the energy spectra shown in Fig.~\ref{fig:spectra}  exhibit a continuous component rising at the lowest energies. This increase in event rate below 500~eVee is attributed to the dominating neutron induced nuclear recoil background, as shown in~\cite{Ricochet:2022pzj}. Finally, the steep rise below 250~eVee, mostly visible for RED227 and RED237 which have lower resolutions than RED177, is due to noise induced triggers.

\begin{table}
\begin{center}
\begin{tabular}{cccc} 
 \hline
 Detector & Baseline & 1.30 keV & 10.37 keV \\ 
 \hline
 RED177 & $30.8\pm0.1$ & $39.2\pm0.6$ & $75.6\pm0.3$ \\ 
 \hline
 RED227 & $33.9\pm0.1$ & $40.7\pm1.1$ & $76.7\pm0.4$ \\
 \hline
 RED237 & $37.3\pm0.2$ & $43.4\pm2.1$ & $74.6\pm0.9$ \\
 \hline
\end{tabular}
\caption{Observed baseline and peak resolutions (RMS) at the  1.30~keV and 10.37~keV lines for the three planar detectors.}
\label{tab:fano}
\end{center}
\end{table}

\begin{figure*}[t]
    \centering
    \includegraphics[width=.49\textwidth]{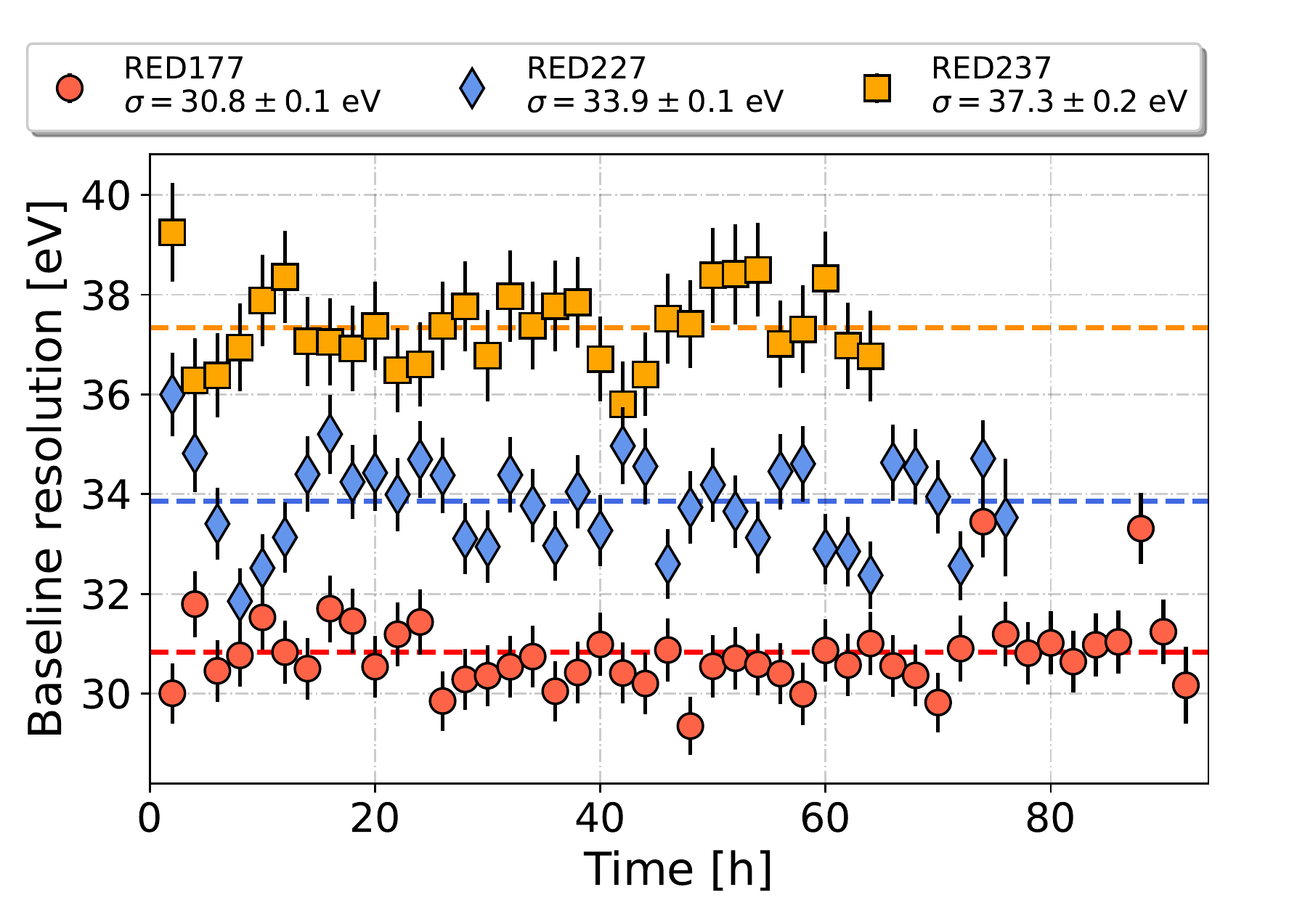}
    \includegraphics[width=.49\textwidth]{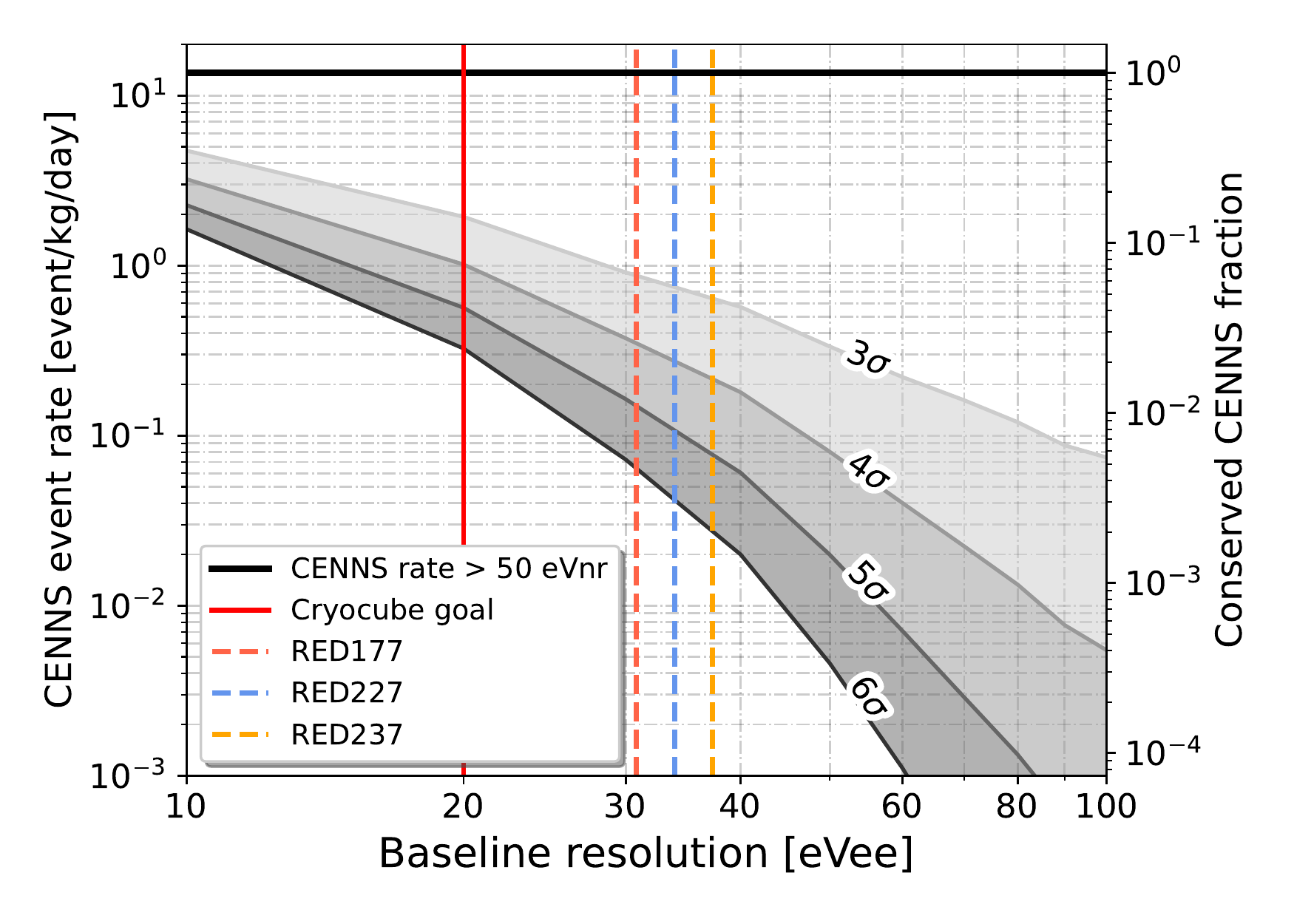}
    \caption{{\bf Left:} Time evolution of the observed baseline energy resolution for RED177, RED227, and RED237 during separate continuous data acquisitions of 92~h, 75~h, and 65~h, respectively, leading to the energy spectra shown in Fig.~\ref{fig:spectra}. {\bf Right:} CENNS event rate (left y-axis) and CENNS signal acceptance (right y-axis) as a function of the baseline ionization energy resolution assuming a 3-to-6~$\sigma$ ionization threshold. The ionization yield model considered here is the standard Lindhard quenching factor on germanium with $k=0.157$. The performance of the three detectors presented in this work are shown in colored dashed lines while the CryoCube targeted resolution is represented by a solid red line.}
    \label{fig:TimeHO}
\end{figure*}

 Figure~\ref{fig:TimeHO} (left panel) presents the time variation of the observed baseline energy resolution of the three detectors over their continuous data acquisition periods. Each data point corresponds to an averaged resolution value over two hours. The energy resolution of all three detectors was extremely stable over the entire acquisition time, with a dispersion of less than 1~eVee, and with no apparent time degradation. This result is particularly important in the context of the \Ricochet{} experiment aiming at measuring the CENNS process with a high precision with year-long data acquisition time. The time averaged baseline energy resolution (RMS) of the detectors was found to be  $30.8 \pm 0.1$~eVee (RED177), $33.9 \pm 0.1$~eVee (RED227), and $37.3 \pm 0.2$~eVee (RED237), see Table~\ref{tab:fano}. This demonstrates for the first time that a 30~eVee ionization resolution has been achieved in a cryogenic bolometer operated at 17~mK (RED177), and that all three detectors in the array reached resolutions on the 30~eVee-scale. This also equals the performance of the best PPC ionization-only Ge detectors \cite{CDEX:2017hbo,nGeN:2022uje,CONUS:2020skt,Collar:2021fcl}, operated at 77~K and thus with much relaxed heat load constraints. The worse energy resolution of RED237 is likely due to additional Johnson noise from its higher impedance constantan cabling, see Sec.~\ref{sec:setup}, which reduced the gain of the RED237 HEMT preamplifier by a factor of two compared to the other detectors RED137 and RED227, further degrading its signal-to-noise ratio.

 \section{Mitigating the non-ionizing low-energy excess}
 \label{sec:HO}

 The choice of a dual heat-and-ionization readout for the CryoCube was initially motivated to reject the gamma induced electronic recoil and beta-decay related surface backgrounds~\cite{RICOCHET:2021gkf}. However, as the heat energy threshold of the bolometers improved, a new and overwhelming source of background of unknown origin has been identified. The latter, observed in all cryogenic experiments with a threshold below 100~eV, is the  subject of ongoing intense worldwide
investigations (see~\cite{Fuss:2022fxe} for a detailed review). From the combined \Ricochet{}-CryoCube~\cite{RICOCHET:2021gkf} and EDELWEISS~\cite{EDELWEISS:2019vjv,EDELWEISS:2022ktt} observations, this low-energy excess has been found to be non-ionizing. In~\cite{EDELWEISS:2022ktt} an upper limit on its ionization yield of $<4\times10^{-4}$ (at 90\% C.L.) has been derived. The rate of this excess, also referred to  as ``heat only'', has been measured to be between $10^6$ and $10^9$ events/day/kg/keV below 100~eV in different low-threshold cryogenic experiments~\cite{Fuss:2022fxe}. As such, this background is about 5 orders of magnitude larger than the CENNS signal expected with the future \Ricochet{} experiment at ILL~\cite{Ricochet:2022pzj}. If not mitigated efficiently, this ``heat only'' background could therefore jeopardize the CENNS sensitivity of the future \Ricochet{} experiment.

This background can be rejected by requiring the presence of an ionization signal above a certain threshold. However, this comes with a loss in efficiency on the CENNS signal that depends on the ionization yield of nuclear recoils. For instance, a 3$\sigma$ to 6$\sigma$ ionization threshold cut leads to a rejection power ranging from about $10^3$ to $10^9$~\cite{Workman:2022ynf}. Figure~\ref{fig:TimeHO} (right panel) shows the remaining CENNS event rate passing the ionization threshold (left y-axis), and the corresponding CENNS survival probability (right y-axis), as a function of the ionization baseline resolution. The calculations were done for rejection levels ranging from 3$\sigma$ to 6$\sigma$ and considering the standard Lindhard quenching factor model with $k=0.157$~\cite{osti_4536390}. Our results suggest that in the case of an overwhelming ``heat only'' excess, as currently observed in all cryogenic experiments, reducing the baseline ionization resolution is pivotal to the success of the future \Ricochet{} experiment. Indeed, assuming a 4$\sigma$ ionization threshold corresponding to a $10^5$ rejection factor, as would be required from our observed ``heat only'' rate~\cite{EDELWEISS:2019vjv,RICOCHET:2021gkf,EDELWEISS:2022ktt}, the CENNS event rate increases from $7\times10^{-3}$ to 1~event/kg/day by reducing the baseline energy resolution from 91~eVee, as achieved in~\cite{Phipps:2016gdx}, to our targeted 20~eVee. With the already demonstrated baseline ionization resolutions presented in this work, the expected CENNS event rate surviving a 4$\sigma$ ionization threshold cut would be between 0.4 (RED177) and 0.2 (RED237) events/kg/day. Note that we obtain similar results when considering the recently measured CONUS quenching factor~\cite{Bonhomme:2022lcz}, and a significantly enhanced CENNS event rate by a factor of 3 when considering instead the measurement from~\cite{Collar:2021fcl}. The strong dependence of the \Ricochet{} experiment's sensitivity to the assumed nuclear recoil ionization yield highlights the need for precise measurements of this value near 100~eV recoil energy~\cite{CRAB:2022rcm,Albakry:2023xrd}, a task that should be facilitated with the major improvements in charge resolution achieved in this work.

\section{Conclusion and Outlook}
\label{sec:conclusion}

In this paper we have presented the first ionization readout performance from three \Ricochet{} germanium detector prototypes with their dedicated HEMT-based preamplifiers mounted in a CryoCube sub-element. All three detectors demonstrated 30~eVee-scale ionization energy resolutions. This is comparable to the best resolutions achieved with ionization-only Ge detectors that operate with much less stringent heat load constraints~\cite{CDEX:2017hbo,nGeN:2022uje,Collar:2021fcl,CONUS:2020skt}. This result corresponds to an improvement by a factor 7 and 11 with respect to previously achieved ionization resolutions from the EDELWEISS~\cite{EDELWEISS:2016nzl} and SuperCDMS~\cite{SuperCDMS:2014cds} collaborations, both using cryogenic Ge detectors operated at 20 and 50~mK, respectively. It is also a factor of 3 better than the previously achieved 91~eVee resolution from the first cryogenic HEMT-based preamplifier~\cite{Phipps:2016gdx}. Additionally,  our HEMT-based preamplifier dissipates about ten times less power per channel compared to \cite{Phipps:2016gdx}, opening the possibility to readout 150 channels with heat sinking the cold electronics at 1~K, as required from the CryoCube detector assembly specifications.

In Sec.~\ref{sec:HO}, we investigated the implication of reaching such low ionization resolutions on the mitigation of the limiting low-energy excess (so-called ``heat only''). We found that by imposing an ionization threshold cut with our targeted 20~eVee ionization resolution, this overwhelming background could be reduced by about $10^5$ while retaining a CENNS signal event rate of 1 event/kg/day. This study also highlighted the crucial need for the \Ricochet{} experiment to improve by about an order of magnitude the ionization performance previously achieved by the EDELWEISS and SuperCDMS collaborations to secure its CENNS sensitivity.

Though the HEMT-based common source preamplifiers considered here allowed us to first study the HEMT properties and achieve a 30~eVee-scale ionization resolution, they do not fully comply with the CryoCube readout specifications. Due to their low-gain ($\sim$10) and their required low-noise amplifiers at room temperature, the collaboration has developed optimized HEMT-based preamplifiers that will be tested in the coming months~\cite{Ricochet:2021cqv}. The increased gain of these preamplifiers will lead to a lower sensitivity to environmental noise with a gain of 100. Additionally, they will be compatible with our newly developed room-temperature amplifiers and digitizer which will allow us to read out the heat and ionization channels of these detectors simultaneously. Lastly, using this dedicated CryoCube readout electronics together with reduced parasitic capacitance flexible PCB, currently under development, we expect to further improve our ionization resolution down to the targeted 20~eVee. 

Such ionization performance, combined with already demonstrated 20~eV-scale baseline heat energy resolution~\cite{EDELWEISS:2019vjv}, will allow to reduce the particle identification threshold down to the 100~eVnr-scale (eV nuclear recoil), hence one order of magnitude lower than that achieved in current dual-readout cryogenic dark matter experiments such as EDELWEISS~\cite{Hehn:2016nll}, SuperCDMS~\cite{SuperCDMS:2014cds}, and CRESST~\cite{CRESST:2019jnq}. Finally, this will simultaneously allow for 1) a precise CENNS measurement with \Ricochet{}, 2) the possibility to directly measure the nuclear recoil ionization yield down to 100~eVnr, and 3) further characterizations and understanding the origin of the overwhelming low-energy excess affecting all low-threshold cryogenic experiments, drastically limiting their CENNS and low-mass dark matter sensitivities~\cite{Fuss:2022fxe}.

\section*{Acknowledgments}

This project received funding from the European Research Council (ERC) under the European Union’s Horizon 2020 research and innovation program under Grant Agreement ERC-StG-CENNS 803079, the French National Research Agency (ANR) within the project ANR-20-CE31-0006, the LabEx Lyon Institute of Origins (ANR-10-LABX-0066) of the Université de Lyon, within the Plan France2030, and the NSF under Grant PHY-2209585 and PHY-2013203. We are grateful to Laurent Couraud and Antonella Cavanna for the HEMT production and development which is supported in part by the French network RENATECH. A portion of the work carried out at MIT was supported by DOE QuantISED award DE-SC0020181 and the Heising-Simons Foundation. This work is also partly supported by the Ministry of Science and Higher Education of the Russian Federation.

\bibliographystyle{JHEP}
\bibliography{Refs}

\providecommand{\href}[2]{#2}\begingroup\raggedright\begin{thebibliography}{10}

\bibitem{Akimov:2017ade}
{\bf COHERENT} Collaboration, D.~Akimov et~al., {\it {Observation of Coherent
  Elastic Neutrino-Nucleus Scattering}},  {\em Science} {\bf 357} (2017),
  no.~6356 1123--1126, [\href{http://arxiv.org/abs/1708.01294}{{\tt
  arXiv:1708.01294}}].

\bibitem{COHERENT:2020iec}
{\bf COHERENT} Collaboration, D.~Akimov et~al., {\it {First Measurement of
  Coherent Elastic Neutrino-Nucleus Scattering on Argon}},  {\em Phys. Rev.
  Lett.} {\bf 126} (2021), no.~1 012002,
  [\href{http://arxiv.org/abs/2003.10630}{{\tt arXiv:2003.10630}}].

\bibitem{Ricochet:2022pzj}
{\bf Ricochet} Collaboration, C.~Augier et~al., {\it {Fast neutron background
  characterization of the future Ricochet experiment at the ILL research
  nuclear reactor}},  {\em Eur. Phys. J. C} {\bf 83} (2023), no.~1 20,
  [\href{http://arxiv.org/abs/2208.01760}{{\tt arXiv:2208.01760}}].

\bibitem{Ricochet:2021rjo}
{\bf Ricochet} Collaboration, C.~Augier et~al., {\it {Ricochet Progress and
  Status}},  in {\em {19th International Workshop on Low Temperature
  Detectors}}, 11, 2021.
\newblock \href{http://arxiv.org/abs/2111.06745}{{\tt arXiv:2111.06745}}.

\bibitem{Billard:2018jnl}
J.~Billard, J.~Johnston, and B.~J. Kavanagh, {\it {Prospects for exploring New
  Physics in Coherent Elastic Neutrino-Nucleus Scattering}},  {\em JCAP} {\bf
  11} (2018) 016, [\href{http://arxiv.org/abs/1805.01798}{{\tt
  arXiv:1805.01798}}].

\bibitem{Fuss:2022fxe}
P.~Adari et~al., {\it {EXCESS workshop: Descriptions of rising low-energy
  spectra}},  {\em SciPost Phys. Proc.} {\bf 9} (2022) 001,
  [\href{http://arxiv.org/abs/2202.05097}{{\tt arXiv:2202.05097}}].

\bibitem{Armengaud:2017rzu}
{\bf EDELWEISS} Collaboration, E.~Armengaud et~al., {\it {Performance of the
  EDELWEISS-III experiment for direct dark matter searches}},  {\em JINST} {\bf
  12} (2017), no.~08 P08010, [\href{http://arxiv.org/abs/1706.01070}{{\tt
  arXiv:1706.01070}}].

\bibitem{RICOCHET:2021gkf}
{\bf RICOCHET} Collaboration, T.~Salagnac et~al., {\it {Optimization and
  performance of the CryoCube detector for the future RICOCHET low-energy
  neutrino experiment}},  in {\em {19th International Workshop on Low
  Temperature Detectors}}, 11, 2021.
\newblock \href{http://arxiv.org/abs/2111.12438}{{\tt arXiv:2111.12438}}.

\bibitem{SuperCDMS:2020aus}
{\bf SuperCDMS} Collaboration, I.~Alkhatib et~al., {\it {Light Dark Matter
  Search with a High-Resolution Athermal Phonon Detector Operated Above
  Ground}},  {\em Phys. Rev. Lett.} {\bf 127} (2021) 061801,
  [\href{http://arxiv.org/abs/2007.14289}{{\tt arXiv:2007.14289}}].

\bibitem{EDELWEISS:2019vjv}
{\bf EDELWEISS} Collaboration, E.~Armengaud et~al., {\it {Searching for
  low-mass dark matter particles with a massive Ge bolometer operated
  above-ground}},  {\em Phys. Rev. D} {\bf 99} (2019), no.~8 082003,
  [\href{http://arxiv.org/abs/1901.03588}{{\tt arXiv:1901.03588}}].

\bibitem{Hehn:2016nll}
{\bf EDELWEISS} Collaboration, L.~Hehn et~al., {\it {Improved EDELWEISS-III
  sensitivity for low-mass WIMPs using a profile likelihood approach}},  {\em
  Eur. Phys. J.} {\bf C76} (2016), no.~10 548,
  [\href{http://arxiv.org/abs/1607.03367}{{\tt arXiv:1607.03367}}].

\bibitem{SuperCDMS:2014cds}
{\bf SuperCDMS} Collaboration, R.~Agnese et~al., {\it {Search for Low-Mass
  Weakly Interacting Massive Particles with SuperCDMS}},  {\em Phys. Rev.
  Lett.} {\bf 112} (2014), no.~24 241302,
  [\href{http://arxiv.org/abs/1402.7137}{{\tt arXiv:1402.7137}}].

\bibitem{CDEX:2017hbo}
{\bf CDEX} Collaboration, L.~T. Yang et~al., {\it {Limits on light WIMPs with a
  1 kg-scale germanium detector at 160 eVee physics threshold at the China
  Jinping Underground Laboratory}},  {\em Chin. Phys. C} {\bf 42} (2018), no.~2
  023002, [\href{http://arxiv.org/abs/1710.06650}{{\tt arXiv:1710.06650}}].

\bibitem{nGeN:2022uje}
{\bf \ensuremath{\nu}GeN} Collaboration, I.~Alekseev et~al., {\it {First
  results of the \ensuremath{\nu}GeN experiment on coherent elastic
  neutrino-nucleus scattering}},  {\em Phys. Rev. D} {\bf 106} (2022), no.~5
  L051101, [\href{http://arxiv.org/abs/2205.04305}{{\tt arXiv:2205.04305}}].

\bibitem{Collar:2021fcl}
J.~I. Collar, A.~R.~L. Kavner, and C.~M. Lewis, {\it {Germanium response to
  sub-keV nuclear recoils: a multipronged experimental characterization}},
  {\em Phys. Rev. D} {\bf 103} (2021), no.~12 122003,
  [\href{http://arxiv.org/abs/2102.10089}{{\tt arXiv:2102.10089}}].

\bibitem{CONUS:2020skt}
{\bf CONUS} Collaboration, H.~Bonet et~al., {\it {Constraints on elastic
  neutrino nucleus scattering in the fully coherent regime from the CONUS
  experiment}},  {\em Phys. Rev. Lett.} {\bf 126} (2021), no.~4 041804,
  [\href{http://arxiv.org/abs/2011.00210}{{\tt arXiv:2011.00210}}].

\bibitem{articleHEMT}
Q.~Dong, Y.-X. Liang, D.~Ferry, et~al., {\it Ultra-low noise high electron
  mobility transistors for high-impedance and low-frequency deep cryogenic
  readout electronics},  {\em Appl.~Phys.~Lett.} {\bf 105} (2014)
  013504--013504.

\bibitem{Phipps:2016gdx}
A.~Phipps, A.~Juillard, B.~Sadoulet, B.~Serfass, and Y.~Jin, {\it {A HEMT-Based
  Cryogenic Charge Amplifier with sub-100 eVee Ionization Resolution for
  Massive Semiconductor Dark Matter Detectors}},  {\em Nucl. Instrum. Meth. A}
  {\bf 940} (2019) 181--184, [\href{http://arxiv.org/abs/1611.09712}{{\tt
  arXiv:1611.09712}}].

\bibitem{EDELWEISS:2016nzl}
{\bf EDELWEISS} Collaboration, L.~Hehn et~al., {\it {Improved EDELWEISS-III
  sensitivity for low-mass WIMPs using a profile likelihood approach}},  {\em
  Eur. Phys. J. C} {\bf 76} (2016), no.~10 548,
  [\href{http://arxiv.org/abs/1607.03367}{{\tt arXiv:1607.03367}}].

\bibitem{misiak:tel-03328713}
D.~Misiak, {\em {D{\'e}veloppements de nouveaux d{\'e}tecteurs cryog{\'e}niques
  bas seuils pour la recherche de mati{\`e}re noire l{\'e}g{\`e}re et la
  physique des neutrinos de basse {\'e}nergie}}.
\newblock Theses, {Universit{\'e} de Lyon}, Feb., 2021.

\bibitem{Timetal}
P.~Wikus, S.~Hertel, S.~Leman, et~al., {\it The electrical resistance and
  thermal conductivity of ti 15v-3cr-3sn-3al at cryogenic temperatures},  {\em
  Cryogenics} {\bf 51} (2011), no.~1 41--44.

\bibitem{EDELWEISS:2020fxc}
{\bf EDELWEISS} Collaboration, Q.~Arnaud et~al., {\it {First germanium-based
  constraints on sub-MeV Dark Matter with the EDELWEISS experiment}},  {\em
  Phys. Rev. Lett.} {\bf 125} (2020), no.~14 141301,
  [\href{http://arxiv.org/abs/2003.01046}{{\tt arXiv:2003.01046}}].

\bibitem{Juillard:2019njs}
A.~Juillard et~al., {\it {Low-noise HEMTs for Coherent Elastic Neutrino
  Scattering and Low-Mass Dark Matter Cryogenic Semiconductor Detectors}},
  {\em J. Low Temp. Phys.} {\bf 199} (2019), no.~3-4 798--806,
  [\href{http://arxiv.org/abs/1909.02879}{{\tt arXiv:1909.02879}}].

\bibitem{Ricochet:2021cqv}
{\bf Ricochet} Collaboration, G.~Baulieu et~al., {\it {HEMT-Based 1 K Front-End
  Electronics for the Heat and Ionization Ge CryoCube of the Future Ricochet
  CE$\nu $NS Experiment}},  {\em J. Low Temp. Phys.} {\bf 209} (2022), no.~3-4
  570--580, [\href{http://arxiv.org/abs/2111.10308}{{\tt arXiv:2111.10308}}].

\bibitem{Colas:2021pxr}
{\bf RICOCHET} Collaboration, J.~Colas, J.~Billard, S.~Ferriol, J.~Gascon, and
  T.~Salagnac, {\it {Development of data processing and analysis pipeline for
  the RICOCHET experiment}},  \href{http://arxiv.org/abs/2111.12856}{{\tt
  arXiv:2111.12856}}.

\bibitem{Allen:1999wh}
B.~Allen, W.-s. Hua, and A.~C. Ottewill, {\it {Automatic cross talk removal
  from multichannel data}},  \href{http://arxiv.org/abs/gr-qc/9909083}{{\tt
  gr-qc/9909083}}.

\bibitem{PhysRev.70.44}
U.~Fano, {\it On the theory of ionization yield of radiations in different
  substances},  {\em Phys. Rev.} {\bf 70} (1946) 44--52.

\bibitem{EDELWEISS:2006bcu}
{\bf EDELWEISS} Collaboration, A.~Benoit et~al., {\it {Measurement of the
  response of heat-and-ionization germanium detectors to nuclear recoils}},
  {\em Nucl. Instrum. Meth. A} {\bf 577} (2007) 558--568,
  [\href{http://arxiv.org/abs/astro-ph/0607502}{{\tt astro-ph/0607502}}].

\bibitem{knoll2010radiation}
G.~Knoll, {\em Radiation Detection and Measurement}.
\newblock Wiley, 2010.

\bibitem{LOWE1997354}
B.~Lowe, {\it Measurements of fano factors in silicon and germanium in the
  low-energy x-ray region},  {\em Nucl.~Instrum.~Meth.~A} {\bf 399} (1997),
  no.~2 354--364.

\bibitem{EDELWEISS:2022ktt}
{\bf EDELWEISS} Collaboration, E.~Armengaud et~al., {\it {Search for sub-GeV
  dark matter via the Migdal effect with an EDELWEISS germanium detector with
  NbSi transition-edge sensors}},  {\em Phys. Rev. D} {\bf 106} (2022), no.~6
  062004, [\href{http://arxiv.org/abs/2203.03993}{{\tt arXiv:2203.03993}}].

\bibitem{Workman:2022ynf}
{\bf Particle Data Group} Collaboration, R.~L. Workman et~al., {\it {Review of
  Particle Physics, Sec.~40 Statistics,}},  {\em PTEP} {\bf 2022} (2022)
  083C01.

\bibitem{osti_4536390}
J.~Lindhard, {\it Influence of crystal lattice on motion of energetic charged
  particles},  {\em Kongel. Dan. Vidensk. Selsk., Mat.-Fys. Medd.} {\bf 34}
  (1965), no.~14.

\bibitem{Bonhomme:2022lcz}
A.~Bonhomme et~al., {\it {Direct measurement of the ionization quenching factor
  of nuclear recoils in germanium in the keV energy range}},  {\em Eur. Phys.
  J. C} {\bf 82} (2022), no.~9 815,
  [\href{http://arxiv.org/abs/2202.03754}{{\tt arXiv:2202.03754}}].

\bibitem{CRAB:2022rcm}
{\bf CRAB, NUCLEUS} Collaboration, H.~Abele et~al., {\it {Observation of a
  nuclear recoil peak at the 100 eV scale induced by neutron capture}},
  \href{http://arxiv.org/abs/2211.03631}{{\tt arXiv:2211.03631}}.

\bibitem{Albakry:2023xrd}
M.~F. Albakry et~al., {\it {First measurement of the nuclear-recoil ionization
  yield in silicon at 100 eV}},  \href{http://arxiv.org/abs/2303.02196}{{\tt
  arXiv:2303.02196}}.

\bibitem{CRESST:2019jnq}
{\bf CRESST} Collaboration, A.~H. Abdelhameed et~al., {\it {First results from
  the CRESST-III low-mass dark matter program}},  {\em Phys. Rev. D} {\bf 100}
  (2019), no.~10 102002, [\href{http://arxiv.org/abs/1904.00498}{{\tt
  arXiv:1904.00498}}].

\end{thebibliography}\endgroup

\end{document}